\documentclass[12pt]{article}

\usepackage[dvips,final]{graphicx}
\usepackage{epsfig}
\usepackage{epic}
\usepackage{eepic}
\usepackage{amssymb,amsmath}
\usepackage{times}

\begin{document}

\newcommand{\vect}[1]{{ {\bf #1  }}} 
\newcommand{\uvect}[1]{{ \hat{\bf #1  }} }
\newcommand{\ci}{
	{
		{ {\bf c}}_i
	}
}
\newcommand{\fixme}[1]{

{ \bf{ ***FIXME: #1 }}

}
\newcommand{\half}{\frac{1}{2}}
\newcommand{\cop}{\Omega_i^\sigma}
\newcommand{\copbgk}{{\Omega_i^\sigma}_{\mathrm{BGK}}}
\newcommand{\Dop}{\mathbb D}
\newcommand{\tausig}{\tau^\sigma}
\newcommand{\xsig}{x^\sigma}
\newcommand{\tausigb}{\tau^\bar{\sigma}}
\newcommand{\psisig}{\psi^\sigma}
\newcommand{\psisigb}{\psi^{\bar{\sigma}}}
\newcommand{\nusig}{\nu^\sigma}
\newcommand{\msig}{m^\sigma}
\newcommand{\nsig}{n^\sigma}
\newcommand{\usig}{u^\sigma}
\newcommand{\usiga}{u^\sigma_\alpha}
\newcommand{\Fsig}{F^\sigma}
\newcommand{\Fsiga}{F^\sigma_\alpha}
\newcommand{\upr}{u'}
\newcommand{\upra}{{u'}_\alpha}
\newcommand{\vsig}{v^\sigma}
\newcommand{\vsiga}{v^\sigma_\alpha}
\newcommand{\sumsig}{\sum_\sigma}
\newcommand{\sumsigb}{\sum_{\bar{\sigma}}}
\newcommand{\sumsigsigb}{\sum_{\sigma\bar{\sigma}}}
\newcommand{\sumi}{\sum_i}
\newcommand{\msi}{\msig\sum_i}
\newcommand{\ciao}{c_{i{\alpha_1}}}
\newcommand{\cian}{c_{i{\alpha_n}}}
\newcommand{\cia}{c_{i\alpha}}
\newcommand{\cib}{c_{i\beta}}
\newcommand{\cig}{c_{i\gamma}}
\newcommand{\cid}{c_{i\delta}}
\newcommand{\cs}{c_{\mathrm{s}}}
\newcommand{\rhosig}{\rho^\sigma}
\newcommand{\frt}{\frac{\rhosig}{\tausig}}
\newcommand{\xt}{(\vect{x},t)}
\newcommand{\xpct}{(\vect{x}+\ci,t)}
\newcommand{\xpc}{(\vect{x}+\ci)}
\newcommand{\Ua}{U_\alpha}
\newcommand{\fis}{f_i^\sigma}
\newcommand{\fisb}{\bar{f}_i^\sigma}
\newcommand{\Nis}{N_i^\sigma}
\newcommand{\NiU}{N_i^\sigma({\bf U})}
\newcommand{\Niu}{N_i^\sigma({\bf u})}
\newcommand{\Nivs}{N_i^\sigma({\bf v}^\sigma)}
\newcommand{\Ti}{T_i}
\newcommand{\Ta}{{T}^{(1)}_{\alpha}}
\newcommand{\Tab}{{T}^{(2)}_{\alpha\beta}}
\newcommand{\Tabg}{{T}^{(3)}_{\alpha\beta\gamma}}
\newcommand{\kronab}{\delta_{\alpha\beta}}
\newcommand{\kronag}{\delta_{\alpha\gamma}}
\newcommand{\kronbg}{\delta_{\beta\gamma}}
\newcommand{\oz}[1]{{#1}^{(0)}}
\newcommand{\oo}[1]{{#1}^{(1)}}
\newcommand{\ot}[1]{{#1}^{(2)}}
\newcommand{\ordn}[1]{{#1}^{(n)}}
\newcommand{\partiald}[2]{
	\frac { \partial #1 } { \partial #2 }
}
\newcommand{\partialdd}[2]{
	\frac { \partial^2 #1 } { \partial {#2}^2 }
}

\newcommand{\partialop}[1]{
	\frac { \partial } { \partial #1 }
}
\newcommand{\partialopop}[1]{
	\frac { \partial^2 } { \partial {#1}^2 }
}
\newcommand{\dal}{\partial_\alpha}
\newcommand{\dbe}{\partial_\beta}
\newcommand{\dga}{\partial_\gamma}
\newcommand{\dt}{\partial_{t}}
\newcommand{\dit}{\partial_{1t}}
\newcommand{\dtt}{\partial_{2t}}
\newcommand{\Psa}{\Pi^\sigma_\alpha}
\newcommand{\Psab}{\Pi^\sigma_{\alpha\beta}}
\newcommand{\Psabg}{\Pi^\sigma_{\alpha\beta\gamma}}
\newcommand{\Pa}{\Pi_\alpha}
\newcommand{\Pab}{\Pi_{\alpha\beta}}
\newcommand{\Pabg}{\Pi_{\alpha\beta\gamma}}
\newcommand{\ep}{\epsilon}

\begin{titlepage}

\begin{center}

{\bf \Huge {Steering in computational science: mesoscale modelling and
simulation}}

\end{center}

{\center
{\noindent
J.~Chin, J.~Harting, S.~Jha, 
P.~V.~Coveney\footnotemark[1]  \\\bigskip

{\it \noindent
Centre for Computational Science\\
Christopher Ingold Laboratories\\
University College London\\
20 Gordon Street\\
London WC1H 0AJ\\
U.K.} \\\bigskip

\noindent
A.~R.~Porter, S.~M.~Pickles \\\bigskip
{\it \noindent Manchester Computing\\
Kilburn Building\\
The University of Manchester\\ 
Oxford Road\\
Manchester M13 9PL\\
U.K.} \\
}
}

\footnotetext[1]{
{\tt <P.V.Coveney@ucl.ac.uk>}
}
\center{\today}
\bigskip
\bigskip
\bigskip

\begin{abstract}
This paper outlines the benefits of computational steering for high
performance computing applications. Lattice-Boltzmann mesoscale fluid
simulations of binary and ternary amphiphilic fluids in two and three
dimensions are used to illustrate the substantial improvements which
computational steering offers in terms of resource efficiency and time
to discover new physics. We discuss details of our current steering
implementations and describe their future outlook with the advent of
computational grids.
\end{abstract}

\end{titlepage}

\section{Introduction}
Many phenomena in condensed matter physics operate at length and time
scales which are too large for detailed microscopic modelling. In
microscopic models, based on classical molecular dynamics, usually every
atom or molecule within a system is considered, resulting in rapidly
increasing complexity for increasing problem sizes. This tends to
restrict the currently treatable length scales to the order of several
nanometres and the timescales to the order of nanoseconds since the
computing power required for larger length and timescales is unavailable
today.  For example, a microscopic description of a fluid would track
the position, momentum, and energy of every single fluid molecule. At
room temperature, a microscopic model of a single cubic millimetre of
monatomic gas would have to deal with around 10$^{17}$ variables.
Macroscopic models, on the other hand, usually deal with a smaller
number of variables, often more closely related to physical observables.
For example, a macroscopic model of a fluid would describe its velocity,
density and temperature at various points. 

Statistical mechanics is used to extract macroscopic, thermodynamic
descriptions from the underlying microscopic representation. While
standard methods exist for performing this contraction of description
for systems at thermodynamic equilibrium~\cite{bib:reichl}, we shall be
primarily interested in time-dependent, non-equilibrium systems for
which there is less widespread agreement about their statistical
mechanical description. In addition, comparatively few established
methods exist for the treatment of complex systems which contain
processes operating on several length and time scales.  This situation
presents a general problem for both microscopic and macroscopic models.
Whereas in microscopic descriptions, a vast amount of computational
effort is required to model mesoscopic systems spanning several length
and time scales, macroscopic treatments omit fine details which may give
rise to the characteristic behaviour of the system.

Because of the shortcomings of microscopic and macroscopic descriptions,
and the tremendous importance of fluid dynamics, there is currently
considerable interest in meso\-scale models. These models coarse grain
most of the atomic or molecular details but retain enough of the
essential physics to describe the phenomena of interest. They are
intended to treat systems at intermediate length scales between several
nanometres and a few millimetres, and processes operating on multiple
length and time scales. Examples of such systems can be found
in everyday life: detergents, shampoos, milk, blood, and paint are
materials whose macroscopic behaviour is induced by their microscopic
and/or mesoscopic properties. A very descriptive example can be observed
in the kitchen and can easily be reproduced by the reader: due to the
microscopic interactions between starch molecules in cornflour, as the
starch molecules jam into one another, a mixture of cornflour and water
becomes more difficult to stir if one stirs it quickly. 

Computer modelling forms a valuable tool in understanding the behaviour
of mesoscale
systems~\cite{bib:rivet-boon,bib:succi,bib:rothman-zaleski,bib:gunstensen-rothman-zaleski-zanetti},
and much time and effort is currently being invested in such techniques.
In this article, we provide a brief overview of a few such mesoscale
modelling techniques, and focus in particular on the lattice-Boltzmann
fluid dynamical method. We describe its implementation, and the problems
which arise when the computer implementation is run `statically',
without interaction with either the user or other computational
components.  These problems may be solved by adding extra
functionality which permits the interoperation of the lattice-Boltzmann
code with other programs, such as separate code to monitor and readjust
the simulation in real-time, or a user interface to permit a human to
`steer' the simulation as it runs. We give examples of two such
functionalities: in one, the functionality of the simulation code is
made available, through the use of a wrapper layer, to a high-level
language, which permits versatile control of simulations; in the other,
the code is connected to a general-purpose steering library which
permits users to remotely control many kinds of simulation as they run.

Mesoscale simulations  often require access  to high-end computational
and   visualisation  resources;  we   therefore  proceed   to  discuss
computational  steering in  such a  general  context, and  how it  may
permit much more efficient use of such resources.

The  literature   on  computational  steering   is  quite  substantial
~\cite{Gu:1994}.  However, most papers~\cite{Prins:1999, vetter95} have
focussed  on the  design  and architectural  details of  computational
steering and  then go on to  give a prototype  implementation of their
steering system.   Our paper represents a different  approach, in that
our  motivation  is  a  scientific  problem  for  which  computational
steering is shown to be an  effective tool. We do this by highlighting
the  advantages that  computational steering  brings  over traditional
non-steered simulations.

Computational grids are an increasingly popular  paradigm of
computation, somewhat akin to traditional distributed computing, yet
with a major extension in that they enable  the transparent sharing  and
collective use  of resources (anything from spare PC CPUs to databases
or high-end hardware), which would otherwise  be individual  and
isolated facilities.   Therefore, we discuss  how  the  advent   of
computational  grids  is  expected  to considerably facilitate modelling
with high-performance  computers in general, and computational steering
in particular.

\section{Mesoscale modelling and simulation methods}

\label{Sec:MesoModel} Many mesoscale situations of interest involve fluids,
particularly mixtures of fluids which exhibit complicated behaviour due to the
interactions of their individual molecules. When attempting to model such
systems, one must treat both the bulk flow, or hydrodynamic behaviour, and the
interactions. Hydrodynamic behaviour is very difficult and expensive to treat
by atomistic methods, but relatively straightforward to handle at the continuum
level; conversely, the fluid interactions can be examined at the atomistic
level, but are usually not straightforward to incorporate at the continuum
level. There are some cases where interactions between microscopic particles
may give rise to macroscopic flow behaviour -- for example, Marangoni
flow\cite{bib:scriven-sternling}, where gradients in surface tension (due to,
for example, uneven distribution of surfactant at an interface\cite{bib:grotberg-gaver}) induce
macroscopically observable fluid flow.

Many mesoscale methods start by exploiting the surprising and convenient
fact that it is not necessary to keep track of every single molecule of
a fluid in order to reproduce its hydrodynamic behaviour. Instead, it is
sufficient to group very large numbers of molecules into Lagrangian
`packets of fluid', and treat these packets as self-contained
particles themselves; interactions can take place between these
mesoscopic groupings of particles, rather than their constituent
molecules. Provided that certain restrictions, such as isotropy and
conservation of mass and momentum, are adhered to, then the resulting
large-scale behaviour is extremely similar to, and can often be related
to, that which would result from treating each molecule
individually~\cite{bib:boghosian-coveney,bib:flekkoy-coveney-defabritiis}.

The technique of dissipative particle dynamics
(DPD)~\cite{bib:hoogerbrugge-koelman,bib:espanol-warren} tracks the
position $\vect{r}_i$ and momentum $\vect{p}_i$ of each mesoscopic
particle $i$. The algorithm consists of two stages: in the first,
particle positions are advected according to their momenta, so
$\vect{r}_i \rightarrow \vect{r}_i + \vect{p}_i \delta t$. In the second
stage, the momentum of each particle is updated according to the force
acting upon it, so $\vect{p}_i \rightarrow \vect{p}_i + \sum_{j \neq i}
\vect{F}^C_{ij} + \sum_{j \neq i} \vect{F}^D_{ij} + \sum_{j \neq i}
\vect{F}^R_{ij}$, where $\vect{F}^C_{ij}$ is a conservative interaction
force between different particles, $\vect{F}^D_{ij}$ is a dissipative, viscous
force, and $\vect{F}^R_{ij}$ is a random force to introduce stochastic
fluctuations. An extra force may be introduced to allow interactions
between different particles~\cite{bib:coveney-espanol}.

While DPD permits a continuous range of values for $\vect{r}_i$ and
$\vect{p}_i$, realistic models may be created which discretize the
position, the momentum, or both.  In the Lattice Gas Automaton (LGA)
method~\cite{bib:fhp,bib:wolfram}, mesoscopic particles are only
permitted to occupy points on a Bravais lattice, discretizing space, and
are only permitted to travel along the lattice vectors, discretizing
momentum. The update algorithm also consists of two steps: during the
``advection'' step, particles travel along their velocity vectors to
adjacent lattice sites; in the ``collision'' step, particles at each
individual lattice site undergo collisions during which their velocities
are redistributed in a manner which conserves the total mass and
momentum at each site. Provided the lattice is carefully chosen to
ensure isotropy of the fluid, the large-scale behaviour will be
hydrodynamically correct. A substantial advantage of this algorithm is
that it only requires Boolean operations, and that it is unconditionally
numerically
stable~\cite{bib:frisch-dhumieres-hasslacher-lallemand-pomeau-rivet}.

Fluid mixtures may be simulated by introducing different species onto
the lattice, often denoted by colour: red particles may represent oil,
and blue particles water. Interactions between the different species may
then be introduced by imagining that particles carry a `colour
charge', and experience a force due to the colour field generated by
surrounding particles~\cite{bib:rothman-keller}.

The red and blue fluids may be made immiscible by introducing a force
which compels red particles to travel up the colour-field gradient
towards regions of higher red density, and compels blue particles to
travel towards regions of higher blue density; a mixture of red and blue
particles will then separate into separate single-colour regions.

Porous media may be simulated by blocking off some lattice sites: any
particles which would travel into the blocked sites during the advection
step are bounced back to travel in the opposite direction, producing a
no-slip boundary. The flow of oil-water mixtures in porous media has
been successfully modelled using this
technique~\cite{bib:coveney-maillet-wilson-fowler-almushadani-boghosian}.

An amphiphilic particle, such as a detergent molecule, typically contains two
parts: a water-loving head, and an oil-loving tail. The behaviour of such
molecules can be very complicated. In a mixture of oil and water, such a
particle will seek out regions of oil-water interface, and reduce the
interfacial tension. Solutions of amphiphile in water may spontaneously
assemble to produce a variety of different phases, ranging from simple
spherical or wormlike clusters called ``micelles'', to extensive sponge-like
phases~\cite{bib:gompper-schick}.

The `coloured particle' description may be extended to cover the case
of amphiphiles, by introducing a new species of mesoscopic particle,
which has an orientational degree of freedom. A single such mesoscopic
particle can be regarded as consisting of a red particle and a blue
particle bolted together, so that the whole particle possesses no net
colour charge, but will tend to align itself with the colour field; the
orientation of the mesoscopic particle represents some sort of average
of the orientations of its constituent molecules. Lattice gas models
with amphiphile particles have been used to simulate the effect of
surfactants on oil-water mixtures in porous media~\cite{bib:love-maillet-coveney}, and the
self-assembly of micelles~\cite{bib:boghosian-coveney-emerton,bib:boghosian-coveney-love}.

In the model proposed by Malevanets and
Kapral~\cite{bib:malevanets-kapral}, sometimes called the `Real-coded
Lattice Gas', or `Discrete Simulation Automaton' (DSA) model, the
mesoscopic particles occupy discrete cells in space, but are permitted
to have real-valued velocities.  During the collision step of the
algorithm, the velocities of the particles in each cell are transformed
according to \begin{equation} {\bf v}_i \rightarrow V+\omega[{\bf v}_i
-V], \end{equation} where $\omega$ is a random rotation and $V$ the
centre of mass velocity of particles in a cell. The total momentum in
each cell remains unchanged, producing hydrodynamic behaviour, while the
randomization of the velocities produces dissipative behaviour.
Particles then travel along their velocity vectors to nearby cells.
This model may be generalized to treat immiscible fluid
mixtures~\cite{bib:hashimoto-chen-ohashi} and
amphiphiles~\cite{bib:sakai-chen-ohashi} in much the same manner as with
LGA models.

The lattice Boltzmann (LB) method is a simplification of LGA:
particles have discretized positions and momenta, but rather than
individual Boolean particles being tracked around a lattice, their
real-valued population is stored, resulting in a less noisy method. We
examine LB in more detail in Section \ref{Sec:LatticeBoltzmann}.

\section{Lattice Boltzmann models of immiscible and amphiphilic fluids}
\label{Sec:LatticeBoltzmann}

\newcommand{\fone}{f_1(\vect{r},\vect{v},t)}

The lattice Boltzmann algorithm is a powerful method for simulating
fluid flow. Much of this power lies in the ease with which boundary
conditions can be imposed, and with which the model may be extended to
describe mixtures of interacting complex fluids.  Rather than tracking
the state of individual atoms and molecules, as is done in molecular
dynamics, or tracking individual discrete mesoscopic `packets of
fluid', as in LGA or DSA algorithms, the lattice Boltzmann method
describes the dynamics of the single-particle distribution function of
mesoscopic fluid packets.

\subsection{The Continuum Boltzmann Equation}

In a continuum description, the single-particle distribution function
$\fone$ represents the density of fluid particles with position
$\vect{r}$ and velocity $\vect{v}$ at time $t$, such that the density
and velocity of the macroscopically observable fluid are given by
$\rho(\vect{r},t) = \int \fone {\mathrm d}\vect{v} $ and
$\vect{u}(\vect{r},t) = \int \fone \vect{v} {\mathrm d} \vect{v}$
respectively. In the non-interacting, long mean free path limit, with no
externally applied forces, the evolution of this function is described
by the famous Boltzmann equation,

\begin{equation}
\label{eq:boltzmann}
\left( \dt + \vect{v} \cdot \vect{\nabla} \right) f_1
= \Omega[f_1].
\end{equation}

\noindent The left hand side of the equation describes changes in the
distribution function due to free particle motion; the right hand side
contains the collision operator $\Omega$, describing changes due to
pairwise collisions. Typically, this is an integral expression which can
be hard to work with, so it is often simplified~\cite{bib:bgk} to the
linear Bhatnagar-Gross-Krook, or BGK form:

\begin{equation}
\label{eq:bgk}
\Omega[f] \simeq - \frac 1 \tau \left[ f - f^{\mathrm{(eq)}} \right]
\end{equation}

\noindent The BGK collision operator describes the relaxation, at a rate
controlled by a characteristic time $\tau$, towards a Maxwell-Boltzmann
equilibrium distribution $f^{\mathrm{(eq)}}$. While this is a drastic
simplification, it can be shown that distributions governed by the
Boltzmann-BGK equation conserve mass, momentum, and energy, and obey a
non-equilibrium form of the Second Law of Thermodynamics~\cite{bib:liboff}.
Moreover, it can be shown~\cite{bib:chapman-cowling,bib:liboff} that the
well-known Navier-Stokes equations for macroscopic fluid flow are obeyed on
coarse length and time scales by such distributions.

\subsection{The Lattice Boltzmann Equation}

In a lattice Boltzmann formulation, the single-particle distribution
function is discretized in time and space. The positions $\vect{r}$ on
which $\fone$ is defined are restricted to points $\vect{r}_i$ on a
lattice, and the velocities $\vect{v}$ are restricted to a set
$\ci$ joining points on the lattice; hence, $f_i(\vect{r},t) =
f(\vect{r},\ci,t)$ represents the density of particles at lattice site
$\vect{r}$ travelling with velocity $\ci$, at timestep $t$. The density
and velocity of the simulated fluid are now given by

\begin{equation}
\label{eq:lbe-density}
\rho(\vect{r}) = \sum_i f_i(\vect{r})
\end{equation}

\begin{equation}
\label{eq:lbe-velocity}
\vect{u}(\vect{r}) = \sum_i f_i(\vect{r}) \ci
\end{equation}

\noindent The lattice must be chosen
carefully~\cite{bib:qian-dhumieres-lallemand} to ensure isotropic behaviour of
the simulated fluid.  It can be shown~\cite{bib:he-luo} that the lattice
Boltzmann equation may be rigorously derived by discretizing the continuum
Boltzmann equation; alternatively, it may be regarded as a Boltzmann-level
approximation of its ancestor, the LGA~\cite{bib:mcnamara-zanetti}.

\begin{figure}
\begin{center}
\includegraphics[width=5cm,height=5cm]{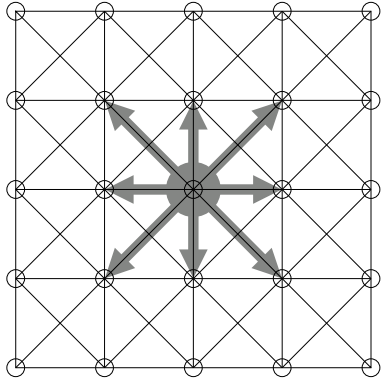}
\end{center}
\caption{
A lattice on which the Boltzmann equation may be discretized. Particles
are only permitted to occupy positions shown by circles.  Particles
occupying the central point are permitted to be either at rest, or to
have one of the eight possible discrete velocities $\ci$ indicated by
the grey arrows.
\label{fig:d2q9-lattice}
}
\end{figure}

The discretized Boltzmann-level description of the fluid may now be
evolved according to a two-step procedure.  In the collision step,
particles at each lattice site are redistributed across the velocity
vectors: this process corresponds to the action of the collision
operator, and usually takes the BGK form:

\begin{equation}
\label{eq:collision}
f_i \leftarrow f_i - \frac 1 \tau
	\left[ f_i - N_i \right],
\end{equation}

\noindent where $N_i =
N_i\left(\rho(\vect{r}),\vect{u}(\vect{r})\right)$ is a
polynomial function of the local density and velocity, which may be
found by discretizing the well-known Maxwell-Boltzmann equilibrium
distribution.

In the advection step, values of the post-collisional distribution
function are propagated to adjacent lattice sites: this corresponds to
particles streaming along their velocity vectors, and is the discretized
equivalent of the left-hand side of the continuum Boltzmann equation:

\begin{equation}
\label{eq:advection}
f_i(\vect{r}+\ci) \leftarrow f_i(\vect{r})
\end{equation}

\noindent Overall, the system obeys the lattice Boltzmann equation (LBE),
 produced by combining the two evolution steps:

\begin{equation}
\label{eq:lbgk}
f_i(\vect{r},t+1) - f_i(\vect{r},t)
= \Omega[f] \\
= - \frac 1 \tau \left[ f_i(\vect{r},t) 
- N_i\left( \rho, \vect{u} \right)\right]
\end{equation}

\noindent It can be shown that the resulting macroscopic density and
velocity fields obey the Navier-Stokes
equations~\cite{bib:chen-chen-matthaeus}.

A well-known drawback of the lattice Boltzmann method is that it is
typically not guaranteed to be numerically stable, and will crash or
produce physically unreasonable results if, for example, the forcing
rate applied to a fluid is too high or if the interparticle interaction
strength is set too high.

\subsection{Multicomponent Interacting lattice Boltzmann Scheme}

\label{Sec:MulticomponentLB} There are several schemes for generalizing
the lattice Boltzmann algorithm to treat multicomponent fluids,
including analogies with
LGA~\cite{bib:gunstensen-rothman-zaleski-zanetti}, imposition of
free-energy functionals~\cite{bib:swift-orlandini-osborn-yeomans},
discretization of a modified form of the continuum Boltzmann
equation~\cite{bib:luo-vdw}, or inclusion of an explicit forcing term in
the collision operator.  The lattice Boltzmann algorithms described in
this paper use the last approach, due to Shan and Chen
~\cite{bib:shan-chen}.

The single-particle distribution function $f_i$ may be extended to the
form $f_i^\sigma$, where each component is denoted by a different value
of the superscript $\sigma$, so that the density and momentum of a
single component $\sigma$ are given by $\rhosig = \sumi \fis$ and
$\rhosig \vect{u}^\sigma = \sumi \fis \ci$ respectively.  The lattice
BGK equation (\ref{eq:lbgk}) now takes the form

\begin{equation}
\label{eq:lbgk-sc}
\fis(\vect{r},t+1) - \fis(\vect{r},t) =
- \frac 1 \tausig
\left[
\fis - N_i(\rhosig, \vect{v}^\sigma)
\right]
\end{equation}

\noindent The velocity $\vect{v}^\sigma$ is found by calculating a
weighted average velocity 


\begin{equation}
\vect{u}' =
\left( \sumsig \frac \rhosig \tausig \vect{u}^\sigma \right)
/ \left( \sumsig \frac \rhosig \tausig \right),
\end{equation}

\noindent and then adding a term to account for external forces,

\begin{equation}
\vect{v}^\sigma = \vect{u}' + \frac \tausig \rhosig \vect{F}^\sigma .
\end{equation}

\noindent The force term $\vect{F}^\sigma$ can take the form  $g
\rho^\sigma \hat{\vect{z}}$ to produce a gravitational force acting in
the $z$-direction. In order to produce nearest-neighbour interactions
between components, it assumes the form

\begin{equation}
\label{eq:colour-colour}
\vect{F}^\sigma = 
- \psisig ( \vect{x} )
\sumsigb g_{\sigma \bar{\sigma}}
	\sumi \psisigb \left( \vect{x} + \ci \right) \ci,
\end{equation}

\noindent where $\psisig ( \vect{x} ) = \psisig ( \rhosig ( \vect{x}))$
is an effective charge for component $\sigma$; $g_{\sigma \bar{\sigma}}$
is a coupling constant controlling the strength of the interaction
between two components $\sigma$ and $\bar{\sigma}$. If $g_{\sigma
\bar{\sigma}}$ is set to zero for $\sigma = \bar{\sigma}$, and a
positive value for $\sigma \neq \bar{\sigma}$ then, in the interface
between bulk regions of each component, particles experience a force in
the direction away from the interface, producing immiscibility. In
two-component systems, it is usually the case that $g_{\sigma
\bar{\sigma}} = g_{\bar{\sigma}\sigma} = g_{br}$.

Amongst other things, this model has been used to simulate spinodal
decomposition~\cite{bib:chin-coveney,bib:gonzalez-nekovee-coveney},
polymer blends~\cite{bib:martys-douglas}, liquid-gas phase
transitions~\cite{bib:shan-chen-liq-gas}, and flow in porous
media~\cite{bib:martys-chen}.

\subsection{Amphiphilic lattice Boltzmann}

As with many other mesoscale fluid methods, amphiphilic fluids may be
treated in the LB framework by introducing a new species of particle
with an orientational degree of
freedom~\cite{bib:chen-boghosian-coveney}. The particles of this species
are each given a vector dipole moment $\vect{d}$ which has maximum
magnitude $d_0$, corresponding to complete alignment of the constituent
molecules. This is represented in the model by a dipole field
$\vect{d}(\vect{x},t)$ representing the average orientation of any
amphiphile present at site $\vect{x}$.  In the advection step, values of
$\vect{d}(\vect{x},t)$ are propagated around the lattice according to

\begin{equation}
\rho^{\mathrm s}(\vect{x},t+1) \vect{d}(\vect{x},t+1)
= \sumi \tilde{f_i^{\mathrm s}}(\vect{x}-\ci,t)
	\tilde{\vect{d}}(\vect{x}-\ci,t),
\end{equation}

\noindent where tildes denote post-collision values. During the
collision step itself, the dipole moments evolve in a BGK process
controlled by a dipole relaxation time $\tau_d$:

\begin{equation}
\tilde{\vect{d}}(\vect{x},t) = \vect{d}(\vect{x},t)
- \frac 1 {\tau_d}
\left[
	\vect{d}(\vect{x},t)-\vect{d}^{\mathrm{(eq)}}(\vect{x},t)
\right].
\end{equation}

\noindent The equilibrium dipole moment $\vect{d}^{\mathrm{(eq)}}$ is
aligned with the colour field $\vect{h}$:

\begin{equation}
\vect{d}^{\mathrm{(eq)}} \simeq \frac {\beta d_0} 3 \vect{h}
\end{equation}

\noindent The colour field contains a component $\vect{h}^c$
due to coloured particles such as oil and water, and a part $\vect{h}^s$
due to dipoles. The former can be found from the populations of
surrounding lattice sites,

\begin{equation}
\vect{h}^c = \sumsig q^\sigma
	\sumi \rhosig(\vect{x}+\ci) \ci,
\end{equation}

\noindent where $q^\sigma$ is a colour charge, such as $+1$ for red
particles, $-1$ for blue particles, and $0$ for amphiphile particles.
The field due to other dipoles turns out to be given by

\begin{equation}
\vect{h}^s (\vect{x},t) = \sumi \left[
\sum_{j \neq 0} f_i^s ( \vect{x} + \ci, t) 
	\vect{\theta}_j \cdot \vect{d}_i (\vect{x}+\vect{c}_j,t)
	+ f_i^s (\vect{x},t) \vect{d}_i(\vect{x},t)
\right],
\end{equation}

\noindent where the second-rank tensor $\vect{\theta}_j$ is defined in
terms of the unit tensor $\vect{I}$ and lattice vector $\vect{c}_j$ as

\begin{equation}
\vect{\theta}_j = \vect{I} - \frac D {c^2} \vect{c}_j \vect{c}_j.
\end{equation}

In the presence of an amphiphilic species, the force on an oil or water
particle includes an additional term $\vect{F}^{\sigma,s}$ to account
for the colour field due to the amphiphiles. By treating an amphiphilic
particle as a pair of oil and water particles with a very small
separation $\vect{d}$, and Taylor-expanding in $\vect{d}$, it can be
shown that this term is given by 

\begin{equation}
\vect{F}^{\sigma,s} \xt  = 
	-2 \psisig\xt g_{\sigma s}
	\sum_{i \neq 0}
		\tilde{\vect{d}}\xpct \cdot \vect{\theta}_i
			\psi^{s} \xpct,
\end{equation}

\noindent where $g_{\sigma s}$ is a constant controlling the strength of
the interaction between amphiphiles and non-amphiphiles.

While they do not possess a net colour charge, the amphiphiles also
experience a force due to the colour field, consisting of a part
$\vect{F}^{s,c}$ due to ordinary species, and a part $\vect{F}^{s,s}$
due to other amphiphiles. These terms are given by

\begin{equation}
\vect{F}^{s,c} = 
	2 \psi^{s} \xt \tilde{\vect{d}}\xt \cdot
	\sumsig g_{\sigma s}
	\sum_{i \neq 0} \vect{\theta}_i \psisig \xpct
\end{equation}

\begin{eqnarray}
\vect{F}^{s,s} &=& 
	- \frac {4D} {c^2}
	g_{ss} \psi^{s}(\vect{x})
	\sumi \left\{ 
		\tilde{\vect{d}}\xpc \cdot \vect{\theta}_i
		\cdot \tilde{\vect{d}}(\vect{x}) \ci
		\right.
		\\
		\nonumber
		&+& \left. \left[
			\tilde{\vect{d}}\xpc
			\tilde{\vect{d}}(\vect{x})
			+ \tilde{\vect{d}}(\vect{x})
			  \tilde{\vect{d}}\xpc
		\right] 
		\cdot \ci
	\right\}
	\psi^{s} \xpc . \\
	\nonumber
\end{eqnarray}

To summarize, the interactions between fluid components are governed by
the coupling constants $g_{br}$, $g_{cs}$, and $g_{ss}$, controlling the
interaction between different sorts of coloured particles, between
coloured particles and amphiphiles, and between the amphiphiles.

While the form of the interactions seems straightforward at a mesoscopic
level, it is essentially phenomenological, and it is not necessarily
easy to relate the interaction scheme or its coupling constants to
either microscopic molecular characteristics, or to macroscopic phase
behaviour. Some theoretical progress has been made in relating LGA
amphiphile models to an underlying microscopic model~\cite{bib:love},
although macroscopic behaviour is very sensitive not only to the values of
the coupling constants, but to the concentrations of each species
present, {\it inter alia}. Different values of these parameters will give rise
to a wide variety of different phases~\cite{bib:gompper-schick}, such as
spherical and wormlike micelles, sponges, lamellae, or droplets:
the phase behaviour can be very difficult to predict beforehand from the
simulation parameters, and brute-force parameter searches are often
resorted to~\cite{bib:boghosian-coveney-love}.

\section{The practicalities of the lattice Boltzmann method}
\label{Sec:Practicalities}

If the sites of a lattice Boltzmann grid are evolved according to the
algorithm described in Section \ref{Sec:LatticeBoltzmann}, then the
state of each site at a given timestep depends only on its state and the
state of the neighbouring cells at the previous timestep, so that LB can
be considered a form of cellular
automaton~\cite{bib:rothman-zaleski,bib:succi}. This spatial locality of
the algorithm translates into memory locality in implementation,
allowing for efficient performance on contemporary commodity computer
architectures which use caching techniques to improve the speed of
memory access, but also enables extremely efficient implementation on
massively parallel computer architectures since, for a lattice split
across CPUs (spatial domain decomposition), only the state of the
lattice sites at the edge of each CPU's chunk of the lattice must be
communicated to other CPUs. A more detailed examination of LB
performance and a comparison with LGA is available in
~\cite{bib:love-nekovee-coveney-chin-gonzalez-martin}.

The following sections describe how two existing lattice Boltzmann codes
were modified to allow for different forms of computational steering.
One code, LB2D, is a light\-weight, single-CPU solver for two-dimensional
problems; steering was not directly bolted on to this code, but instead
a high-level scripting interface was added to allow simulations to be
controlled at runtime, either through high-level scripts, interactive
manipulation, or other processes. The other code, LB3D, is a solver for
three-dimensional problems involving ternary amphiphilic fluids,
designed for use on distributed-memory parallel processing
architectures; steering was added by interfacing the code to a separate
steering library.\footnote{LB3D was recently awarded the gold-star rating
for its excellent scaling properties with large models running on 1024
processors on HPCx, the UK's fastest supercomputer.} 

\subsection{Design of a typical lattice Boltzmann code}

The lattice Boltzmann codes we examine in this article each revolve
around a single data structure, which encapsulates the entire state of a
simulation at a single instant in time.  Specifically, this data
structure contains the complete state of the lattice, with the value of
$f^\sigma_i(\vect{x})$ for all values of $\sigma$, $i$, and $\vect{x}$;
and also a set of simulation parameters.  These parameters may be
divided into two categories: parameters which are static and unchanging,
such as the dimensions of the simulation lattice or parameters
describing the initial state of the system, and parameters which could
conceivably be changed during the course of the simulation, such as
coupling constants and forcing rates.  The code is then structured as a
set of methods which act upon the data in this structure.

These methods can be loosely grouped into categories.  `Constructor' and
`destructor' methods allocate new simulation objects and free the memory
associated with old ones; initialization methods initialize the state of a
simulation object before commencing a given simulation run.  IO methods write
simulation data to disk, and also modify simulation data according to data
saved on disk. These methods can load and save complete simulation states to
and from disk, as well as loading, for example, porous media data, and saving
information such as the fluid density field to disk. The save-data methods
consist of two parts: one generic routine which produces a block of data
corresponding to a physical field (such as density or pressure), and another
routine which saves this block of data to disk in a specific format (such as
raw binary, a portable binary format known as XDR\cite{bib:rfc1832}, or a
portable image format known as PNG).

Evolution methods perform advection or collision processes on a
simulation object, and generally consume the majority of the CPU time in
a given simulation run.  Finally, boundary condition methods alter the
lattice, for example, to maintain a constant fluid density or
composition at the edges of the simulated region.

\subsection{Traditional simulation methodology and its drawbacks}

Traditionally, large, compute-intensive simulations are run
non-interactively. A text file describing the initial conditions and
parameters for the course of a simulation is prepared, and then the
simulation is submitted to a batch queue, to wait until there are enough
resources available to run the simulation. The simulation runs entirely
according to the prepared input file, and outputs the results to disk
for the user to examine later.

This mode of working is sufficient for many simple investigations of
mesoscale fluid behaviour; however, it has several drawbacks.  Firstly,
consider the situation where one wishes to examine the dynamics of the
separation of two immiscible fluids: this is a subject which has been of
considerable interest in the modelling community in recent
years~\cite{bib:gonzalez-nekovee-coveney,bib:kendon-cates-pagonabarraga-desplat-bladon}.
Typically, a guess is made as to how long the simulation must run before
producing a phase separation, and then the code is run for a fixed
number of timesteps. If a phase transition does not occur within this
number of timesteps, then the job must be resubmitted to the batch
queue, and restarted. However, if a phase transition occurs 
in the early stages of the simulation, then the rest of the compute time
will be spent simulating an equilibrium system of very little interest;
worse, if the initial parameters of the system turn out not to produce a
phase separation, then all of the CPU time invested in the simulation
will have gone to waste.

Secondly, the input file often takes the simple form of a list of
parameters and their values, but this may not be sufficiently expressive
to describe the boundary conditions one may wish to apply, or the
conditions under which they are to be applied.  For example, to simulate
the flow of a fluid mixture through a porous medium, it is necessary to
equilibrate the flow of a single component through the medium first,
before introducing the fluid mixture, in order to prevent the behaviour
of the mixture from being affected by transients present as the flow
field develops.

\subsection{High-level control of simulation codes: scripting}

For every new and complicated boundary condition one wishes to impose,
it is in principle possible to write a corresponding new subroutine in
the simulation code, add an option in the input file to switch this
boundary condition on or off, and recompile the simulation code.
However, in practice this leads to redundancy and overcomplication, or
``bloat'',  in the simulation code, and also to excessively complicated
or verbose input files. Bloated code will, in the long term, become
difficult to maintain or change, and more complicated input file syntax
makes it harder for new users to learn how to use the code.

An alternative strategy is to abandon the concept of an input file
altogether, and instead to control the simulation from a script written
in a high-level language. This has several advantages.

Firstly, provided that enough access to the simulation data structures
is provided to the scripting layer, new boundary conditions may be
formulated, tested, and run with ease, without requiring the code to
be recompiled.  The core of the number-crunching code stays small and
maintainable as a result.

Secondly, a high-level language provides conditional and loop
structures, so simulations may be given much more detailed instructions
than simply to run for a fixed number of timesteps: for example, a
simulation of fluid phase separation could be instructed to run until
the fluid components have separated to a certain degree, and to then
stop.

Thirdly, writing the core of the simulation code in a language like C or
Fortran but controlling its behaviour through a higher-level language
allows the programmer to easily interface the simulation code with other
components (such as image generation libraries) via the high-level
language, which avoids the necessity of dealing with tedious low-level
details of interfacing to many third-party libraries. This strategy also
avoids incurring the performance penalty that would result from writing
the entire simulation code in a higher-level language.

The approach of making a piece of code such as the simulation solver
available as a self-contained reusable object to some higher-level
``glue'' layer is often termed ``componentization''. In this case, the
glue layer is the high-level language; in the more general case, it
could be a Grid fabric layer such as Web Services, allowing
interoperation across the network of components running on different
machines.

The high-level language chosen to script LB2D was Perl, a powerful
language popular, amongst other things, for its ability to interface
with, or `glue', external components, and also for the wide variety of
freely-available Perl code which can be easily accessed from scripts
written in the language~\cite{bib:cpan}. Constructing a functional Perl
interface to the simulation code required little more than writing a
formal description of the C subroutines comprising the simulation
code~\cite{bib:jenness-cozens}; interfacing code to other popular
scripting languages such as TCL, Python, or Ruby is typically just as
easy.

\subsection{Parameter space exploration using high-level scripting facility}

For an algorithm which runs the risk of encountering numerical
instabilities, it is desirable to know the regions of parameter space in
which one can operate without expecting to encounter such problems: for
example, when studying the behaviour of an interface between two fluid
components, it is useful to know how high the surface tension can be set
before numerical instabilities are introduced, resulting in a simulation
crash.

A crude approach to map out this region is to guess the size and
location of a region of parameter space which will contain a region of
stability, and blindly launch many simulations over this space.

A slightly more versatile approach made possible by scripting is to
start with a known-stable point in parameter space, and an initial
direction in parameter space. A simulation is started at the
known-stable point, and if it completes successfully, another one is
started at an adjacent point in parameter space, until eventually a
numerically unstable regime is found; successive simulations can then be
launched to home in on the location of the stable/unstable boundary.

\begin{figure}
\begin{center}
\includegraphics{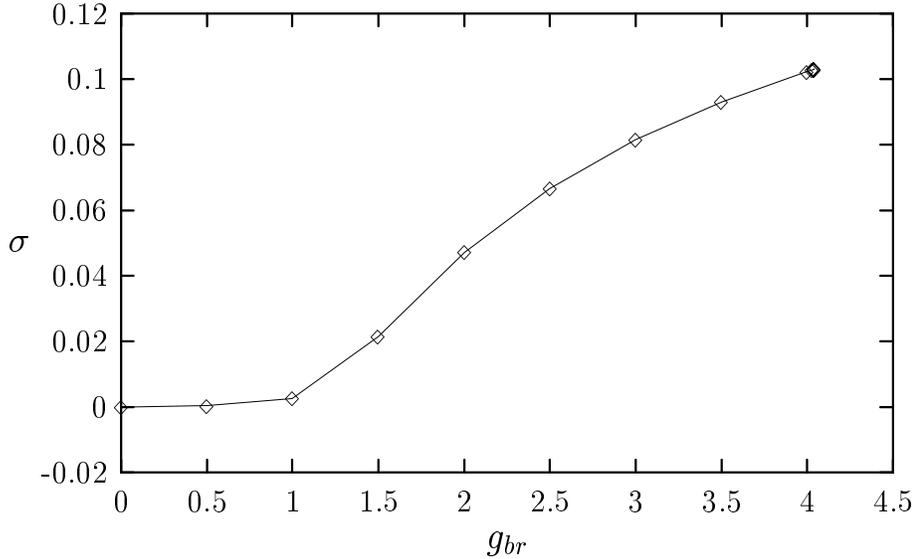}
\end{center}
\caption{
Graph of the surface tension $\sigma$ (in lattice units) of an interface
between two immiscible fluids, against the value of the coupling
constant $g_{br}$. The high-$g_{br}$ limit of the graph indicates the
maximum value, beyond which simulations will become unstable.
\label{fig:steersigma}
}
\end{figure}

In the Shan-Chen LB model, the interaction force between components,
described in equation (\ref{eq:colour-colour}), gives rise to a surface
tension at the interface between regions of different components; the
strength of this interaction, and therefore the magnitude of the surface
tension, is controlled by the coupling constant $g_{br}$. 

A simple parameter-space investigation is to take an interface between
two immiscible fluids, and run simulations with increasing $g_{br}$, and
therefore increasing surface tension, until numerical instability sets
in. The script controlling this process ensures that each simulation
runs for long enough that the system reaches equilibrium. If a
simulation succeeds, then the surface tension is raised; if not, then it
is lowered, and the boundary is located using an interval bisection
algorithm. The results of such an investigation are shown in Figure
\ref{fig:steersigma}, in which the surface tension was calculated for
each value of $g_{br}$ for which a simulation could successfully be run.

\subsection{Scripted boundary conditions}

\newcommand{\dropw}{1cm}
\newcommand{\droph}{4cm}
\begin{figure}

\begin{center}

	\mbox{
		\mbox{
			\begin{tabular}{c}
				\includegraphics[width=\dropw,height=\droph]
					{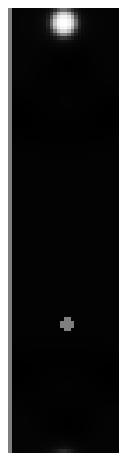} 
				\\
				{Timestep: 50}
				\\
			\end{tabular} 
		} 
		\mbox{
			\begin{tabular}{c}
			\includegraphics[width=\dropw,height=\droph]
				{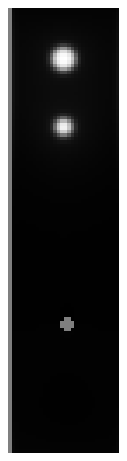}
				\\
				{1000}
				\\
			\end{tabular} 
		}
		\mbox{
			\begin{tabular}{c}
			\includegraphics[width=\dropw,height=\droph]
				{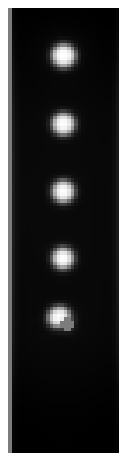}
				\\
				{2650}
				\\
			\end{tabular} 
		}
		\mbox{
			\begin{tabular}{c}
			\includegraphics[width=\dropw,height=\droph]
				{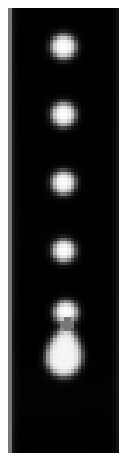}
				\\
				{4450}
				\\
			\end{tabular} 
		}
		\mbox{
			\begin{tabular}{c}
			\includegraphics[width=\dropw,height=\droph]
				{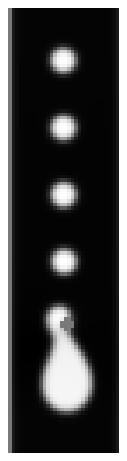}
				\\
				{7250}
				\\
			\end{tabular} 
		}
		\mbox{
			\begin{tabular}{c}
			\includegraphics[width=\dropw,height=\droph]
				{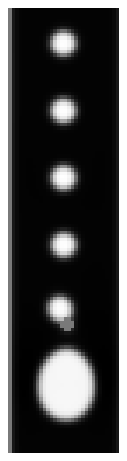}
				\\
				{7850}
				\\
			\end{tabular} 
		}
	}

\end{center}

\caption{ \label{fig:droplets}
Simulation of droplet coalescence on an obstacle during channel flow
using scripted boundary conditions
}
\end{figure}

The use of scripting allows extremely versatile and dynamic
specification of boundary conditions in a simulation. Consider a stream
of droplets flowing through a channel until they meet an obstacle: the
droplets accumulate on the obstacle and coalesce to form a larger
droplet, until the resulting droplet becomes too large and breaks free
to travel further down the channel.  Conventionally, one might
investigate this situation by first running a simulation of
single-component flow through the channel with the obstacle, until the
velocity field equilibrates, and then starting a new simulation from
this one in which droplets are added to the channel, either through
manual intervention, or by writing an additional piece of code which
periodically generates a new droplet near the entrance to the channel.

However, it is also possible to perform the simulation in one single
run, using a script. This script equilibrates the single-component
velocity field, and then automatically introduces a droplet to the
channel entrance. It then waits, monitoring the simulation state, until
the droplet has moved sufficiently far down the channel that a new one
may be introduced without colliding with it. The results of such a
simulation are shown in Figure \ref{fig:droplets}: the
automatically-generated droplet stream is induced by the obstacle to
coalesce into a single large droplet, which then breaks off shortly
before timestep 7850.  The advantage of this approach is that it does
not require human intervention to restart the simulation after the
velocity field has equilibrated, nor does it require a fixed droplet
generation rate to be set: if such a rate were set slightly too high,
then droplets may collide with one another before reaching the obstacle,
a situation avoided by the use of a dynamically-specified boundary
condition.

\section{Steering lattice-Boltzmann simulations through a generic library interface}
\label{Sec:Implementation}
In this section, we present a discussion of the way in which we have
implemented computational steering for LB3D within the ongoing RealityGrid
project \cite{bib:ReG}. The RealityGrid project aims to enable the
modelling and simulation of complex condensed matter structures at the
molecular and mesoscale levels as well as the discovery of new materials
using computational grids. The project also involves biomolecular
applications and its long term ambition is to provide generic
computational grid based technology for scientific, medical and commercial
activities.

\subsection{Motivation}
\label{Sec:Imp:Motivation} 
Within RealityGrid, the way in which computational steering is implemented is
driven by a desire to enable existing scientific computer programs (often
written in Fortran90 and designed for multi-processor/parallel supercomputers)
to be made steerable while minimising the amount of work required. Minimising
the number of changes that a scientist must make to an existing program is
important since it encourages him to take responsibility for this work.
Consequently, the scientist understands the changes that are required and can
continue to maintain the software in the future.

In the light of these requirements, we chose to implement the steering
software as a library written in C and thus callable from a variety
of languages (including C, C++ and Fortran90). The library completely
insulates the application from any implementation details. For
instance, the process by which messages are transferred between the
steering client and the application (e.g. via files or
sockets) is completely hidden from the application code.

Different scientists favour various 
techniques for writing programs intended to run on the specialist
architectures of large supercomputers. Our steering library
therefore does not assume or prescribe any particular
parallel-programming paradigm (e.g. message passing or shared
memory).

Obviously, a scientist does not want a failure in the steering system
(such as a loss of connection to the application) to result in
valuable computing time being lost. We have therefore designed the
steering protocol so that, insofar as is possible, the steering is
never made critical to the simulation process. The protocol enables a
steering client to attach and detach from a running application
without affecting its state.

The scientist's ability to monitor the state of his simulation and
use this to inform his steering decisions plays a key role in
computational steering.  While a steering client provides some information
via the simulation's monitored parameters, a visualisation of some
aspect of the simulation's state is often required. In our
architecture this visualisation is created by a second software
component.

\subsection{Requirements}
\label{Sec:Imp:Requirements} 
In order to make use of the steering library, an application must
satisfy certain requirements. In particular, the application must have
a logical structure such that there exists a point (which we term a
\emph{breakpoint}) within a control loop at which it is possible to
carry out the following steering tasks:

\renewcommand{\labelenumi}{\roman{enumi})}
\begin{enumerate}
\item emit a consistent representation of the state of the 
application's steerable and monitored parameters;
\item accept a change to one or more steered parameters;
\item emit a consistent representation (data sample) of part of the 
      system being simulated (e.g. for visualisation);
\item take a checkpoint or restart from an existing checkpoint.
\end{enumerate}

\renewcommand{\labelenumi}{\arabic{enumi}.}

While all of these things must, theoretically, be possible at the
breakpoint, it is up to the scientist as to how many of them his
application actually supports. For instance, enabling the application
to restart from a checkpoint during execution might be a
difficult task and therefore need only be attempted if the scientist
particularly wants the functionality that that facility will bring.

\subsection{The steering library}
\label{Sec:Imp:SeeringLib} 
The steering library itself consists of two parts: an application side
and a client side.  The client side is intended to be used in the
construction of a steering client.  We have developed a generic
steering client using C++ and Qt (a GUI toolkit)~\cite{Trolltech:QT} which is capable of
steering any application that has been `steering enabled' using the
library.

The steering library itself supports a variety of features.  These
include the facility for the application to register both monitored
(read-only) and steerable (changed only through user interaction)
parameters.  Beyond this facility, the library supports a set of
pre-defined commands such as `pause', `resume', `detach' and
`stop.'  In addition to these pre-defined commands, the library also
allows the user to instruct the application to emit or consume any data
sets that it has previously registered.  Similarly, the user may
instruct the application to take a checkpoint or restart from one which
the application has registered.

The latter functionality is particularly important since it provides
the basis of a system that allows the scientist to `rewind' a
simulation (by restarting from a previous checkpoint).  Having done
so, it can then be run again, perhaps after having steered some
parameter or altered the frequency with which data from the simulation
is recorded. The GRASPARC project~\cite{grasparc} is an example of another
system with this functionality.

In order to maximise the flexibility of the library, we use a system
of `reverse communication' with the application.  This means that,
for most actions, the library simply notifies the application of what the latter needs
to do.  It is then the application's responsibility to carry out the
task, possibly using utility routines from the steering library.  This
is consistent with the philosophy mentioned earlier, of allowing the
scientist to decide how much steering functionality he
wishes to implement.

The steering library currently uses files for transmission of the steering
messages. This means either that the application and the steering client must
have access to the same disk or that some other software (known as
`middleware') takes responsibility for transferring the files between specific
locations on the computers running the application and the client. Work in
progress will lift this restriction by introducing direct communication between
the application and client.

\subsection{Computational steering with LB3D}
As noted earlier, our parallel three-dimensional lattice-Boltzmann code
(LB3D) has been interfaced to the RealityGrid steering library, which
allows the user to steer all parameters of the simulation including
coupling constants, fluid densities, relaxation times and even data
dumping frequencies. Steerable data dumping frequencies enable the user to
increase the amount of generated data for parts of the simulation where
the effects of interest are happening. This helps to save an expensive
resource, namely disk space. 

In addition to the features the steering library provides, LB3D has its own
logging and replay facilities which permit the user to `replay' a steered
simulation. This is an important feature since it allows the data from
steered simulations to be reproduced without human intervention. Moreover,
this feature can be used as an `auto-steerer', i.e. multiple simulations
which read different input-files at startup and are `steered' in the same
way can be launched without the need for human intervention during the
simulation. One application of this particular feature is for
studies of how changes in parameters affect a simulation that has evolved for
a given number of timesteps. Another application is the automatic adaptation
of data dumping or checkpointing frequencies. If the user has found from a
manually steered simulation that no effects of interest are expected for a
given number of initial timesteps, he can reduce the amount of data
written to disk for early times of the simulation.

All steered LB3D simulations that are reported in this paper were performed on
64 processors of an SGI Origin 3800 in Manchester, UK. For data visualisation
we used the Visualization Toolkit (VTK)~\cite{bib:vtk} on a workstation in
London. We chose to run the steering client on the same workstation.

\subsubsection{Spinodal decomposition}
\label{Sec:SpinodalDecomp}
As an example of a typical steered simulation with LB3D, let us consider the
miscibility of a binary fluid mixture.  We are interested in the behaviour of
the system for different values of the coupling constant $g_{br}$ which
controls the strength of the interaction between both fluids, which we call
`blue' and `red' here (see equation (\ref{eq:colour-colour})). By
interacting with a single ongoing simulation, we can change $g_{br}$ `on the
fly' and immediately see how the fluid mixture behaves. Depending on the
phenomena we are interested in, we can `steer' the fluid into miscible or
immiscible states. This technique can as well be utilised to find optimal
values of $g_{br}$ to study spinodal decomposition. Spinodal decomposition
takes place if an incompressible binary fluid mixture is forced into
thermodynamically unstable regions of its phase diagram, i.e. below its
spinodal temperature. In this case, the mixture starts to phase separate into
domains of the two fluids. This effect is important in various industries
because phase separations in products like paints or cosmetics have to be
controlled carefully and many researchers have studied spinodal decomposition in detail~\cite{bib:gonzalez-nekovee-coveney,bib:chin-coveney,bib:coveney-novik,bib:alexander-chen-grunau,bib:grant-elder,bib:furukawa-2d,bib:kendon-cates-pagonabarraga-desplat-bladon,bib:perrot-chan-beysens,bib:rybka-cieplak-salin,bib:solis-olveradelacruz,bib:siggia}.

\begin{figure}
\centerline{\includegraphics[width=15cm]{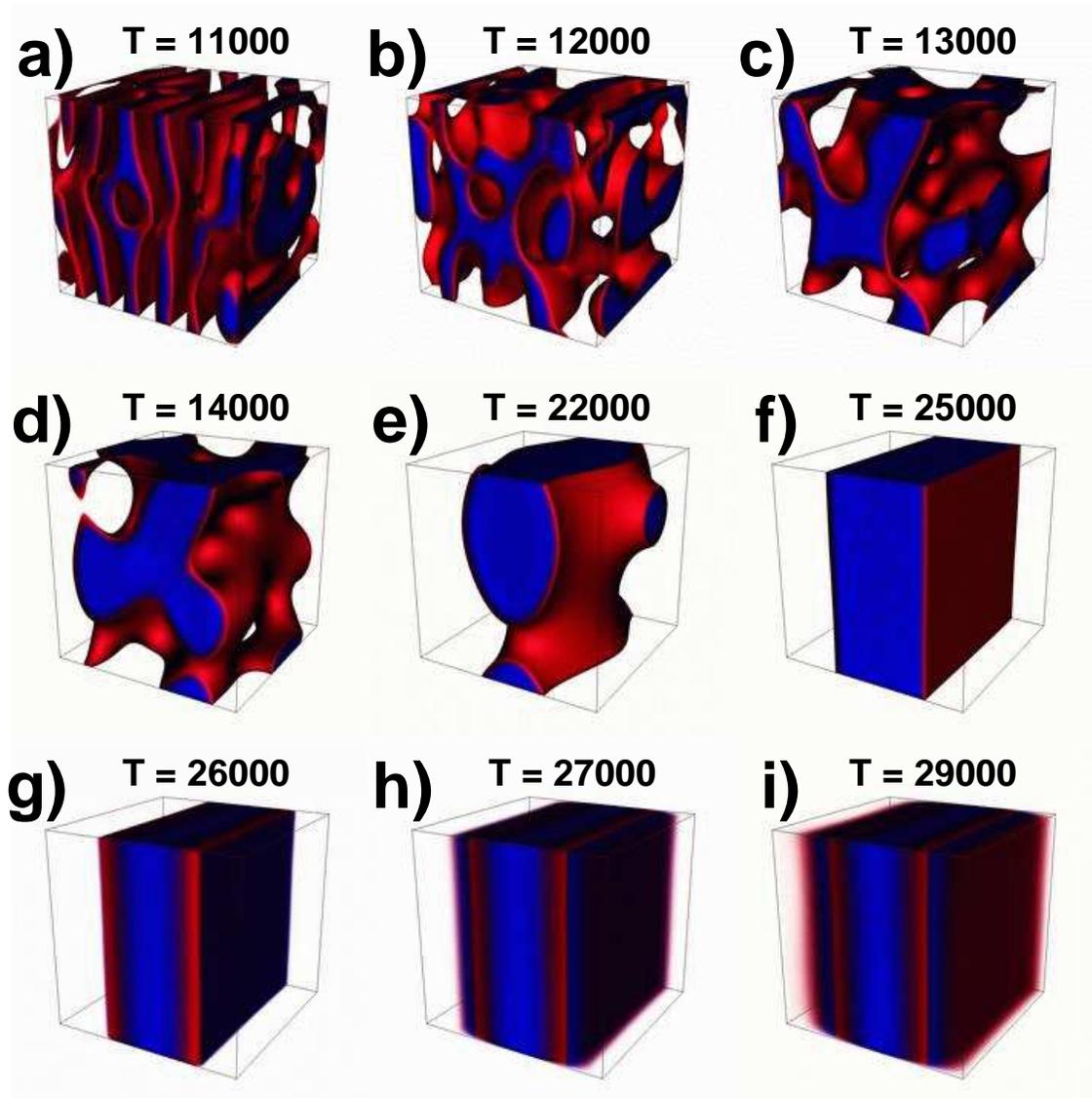}}
\caption{Snapshots of the `colour' field from a steered
lattice-Boltzmann simulation of a 64$^3$ binary fluid mixture of red and
blue particles using the LB3D code. The coupling constant $g_{br}$ between
both fluids is slowly raised from 0.001 to 0.01 during the inital phase of
the simulation so that the phases start to separate until they reach a
fully separated state after 25000 timesteps (a-f). Afterwards, the
coupling is reduced to -0.001 so that the fluids become mixed again (g-i).
In (a-f) the colour red is used to visualise where areas of `red'
dominance start and in  (g-i) the colouring is utilized to show the
diffusion of fluid particles into areas where the other species is
dominant.}
\label{demosteer}
\end{figure}

Figure \ref{demosteer} shows snapshots of volume rendered `colour' fields
(see section \ref{Sec:MesoModel} and \ref{Sec:LatticeBoltzmann} for details).
The colour field describes the net force of different fluid species on a given
lattice site. A value of zero is obtained if forces caused by both fluid types
cancel each other. This takes place at the interface between the fluids. A
`colour' field greater than zero corresponds to areas where the `blue'
fluid dominates while negative values correspond to a domination of `red'
particles respectively. In figures \ref{demosteer}(a-f), areas of `blue'
dominance are rendered in blue and `red'
areas close to the interface are visualised in red. In the remaining
snapshots, the colouring is utilized to visualise the diffusion of one fluid
species into areas where the other species is dominant.

As initial condition for the simulation of the 64$^3$ system, we chose a
mixture, where for both species, each vector on each lattice site is assigned a
random occupation number between zero and 0.7. Relaxation times and masses are
set to unity and the initial value of $g_{br}$ is 0.001. This value is too low
for phase separation to occur. Therefore, $g_{br}$ is slowly raised to a value
of 0.01. Within a few thousand timesteps, both phases start to separate and
after 11000 timesteps of the simulation, a clear structure in the fluid
densities can be observed. It takes until timestep 24000 for the mixture to
reach a fully separated state (figure \ref{demosteer}a-f). 
At timestep 25000 we start to reduce $g_{br}$ again in order to force the
fluids to mix again. The minimum value of the coupling strength used is -0.001.
Soon after reducing $g_{br}$, red fluid particles start to diffuse into areas
of blue dominance and {\em vice versa}. At timestep 29000 the system has
arrived in a nearly fully mixed up state again (figures \ref{demosteer}g-i).

\subsubsection{Parameter searching}
The second example we give to demonstrate the usefulness of computational
steering of three dimensional lattice-Boltzmann simulations is focused on
parameter searches.

As noted previously, our simulations are very resource intensive. Single
simulations might take between hours and days on a large number of processors
of a parallel computer, and storing the generated data requires tens or
hundreds of gigabytes of disk space.  Typical system sizes of our simulations
include 64$^3$, 128$^3$ or 256$^3$ lattices. The data written to disk for a
single measuring timestep of a 256$^3$ lattice requires about one gigabyte of
disk space. For typical simulation lengths of 20000 timesteps and a measuring
frequency of 100 timesteps, 200GB are needed. By reducing the lattice size to
64$^3$, one is able to substantially reduce the amount of data to 3GB.
However, due to the possible occurrence of finite size effects, such small
lattice sizes are often not appropriate. Much of this data might turn out not
to include physically interesting results or the data might be of limited use
because the simulation parameters were not choosen correctly. Moreover, like
other mesoscopic models, the lattice-Boltzmann method contains a number of free
parameters (see section \ref{Sec:LatticeBoltzmann}), resulting in
high-dimensional parameter spaces, although only limited areas may be of
interest. In addition, the phenomena of interest might occur within a limited
time interval in the simulation only. In all these situations, very expensive
compute resources are wasted.

The free parameters of our lattice-Boltzmann ternary amphiphilic fluid
algorithm include the coupling constants between different fluid types (see
section \ref{Sec:MulticomponentLB}). The relationship between these parameters
and experimentally available fluid properties is not well understood.
Therefore, it is important to choose these parameters carefully so as to study
a wide range of phenomena with one or a small number of parameter sets.

Traditionally, such `optimal' parameter sets have been determined by `task
farming' approaches, that is by performing large numbers of small simulations
concurrently, on a large parallel machine or on a large number of small
individual machines~\cite{bib:boghosian-coveney-love}.  This technique allows
one at
least in principle to `scan' the entire parameter space. In practice,
only subspaces can usually be investigated in detail, although these can
be distributed. No human interaction is required after the jobs have been
submitted, which makes it easy to use script based approaches for the
generation of input files and job submission.  However, the available
computing resources are not used very efficiently. Not only is CPU time
wasted in a task farm simulation, the amount of disk space needed to store
the simulation data can be immense. 

For example, we did large scale parameter searches for binary water-surfactant
mixtures. The system size was 64$^3$ and parameters studied were the
surfactant-surfactant coupling constant $g_{ss}$, the surfactant-water coupling
constant $g_{bs}$ (see equation (\ref{eq:colour-colour})), and initial fluid
densities. Masses and relaxation times were kept fixed at unity. We were only
able to study small regions of the available parameter space, i.e. $g_{ss}$ was
varied between -0.001 and -0.006 and values for $g_{bs}$ varied between -0.004
and -0.008. The initial conditions were set as in section
\ref{Sec:SpinodalDecomp}, but the maximum occupation numbers were varied from
0.2 to 0.7 for each fluid individually. In practice, we launched a number of
simulations with different values for $g_{ss}$, $g_{bs}$ and initial densities
and analysed the generated data afterwards. This analysis gave us an idea of
interesting values for the parameters studied and, in principle, on the basis
of these findings one could launch more simulations in order to investigate the
system in more detail.   

However, within a few wall-clock weeks of simulation time, we generated about
300GB of data and used about 30000 CPU hours on an SGI Origin 3800. While the
simulations were performed in a highly automated manner within a couple of
weeks, data analysis has been ongoing for months. Automation of the analysis of
the generated data is much harder because it might be difficult to define the
effects sufficiently well, or impossible to anticipate the effects in advance,
or simply not worthwhile to invest additional effort in the development of 
algorithms to automate the process. 

By considering this example, the disadvantage of conventional parameter searches
is apparent: a significant fraction of the simulations performed in such a
search employ parameters which do not admit interesting phenomena. Nonetheless,
these simulations generate data that has to be analysed. In fact, it is not the
elapsed computing time that makes parameter searches very time consuming, but
extracting information from the data produced.

This analysis time can be significantly reduced by introducing human intuition
into the simulation-analysis loop. The scientist doing the simulations and
analysing the data is usually able to decide whether a parameter set is in a
region of interest long before any given simulation finishes. By providing the
scientist with the possibility to change simulation parameters on the fly, two
goals can be achieved. First, he might be able to `steer' the simulation into
areas of interest: this improves the effective use of CPU cycles and reduces
the amount of produced data. Second, analysing the reduced simulation output
data is much less time consuming.

\begin{figure}
\centerline{\includegraphics[width=15cm]{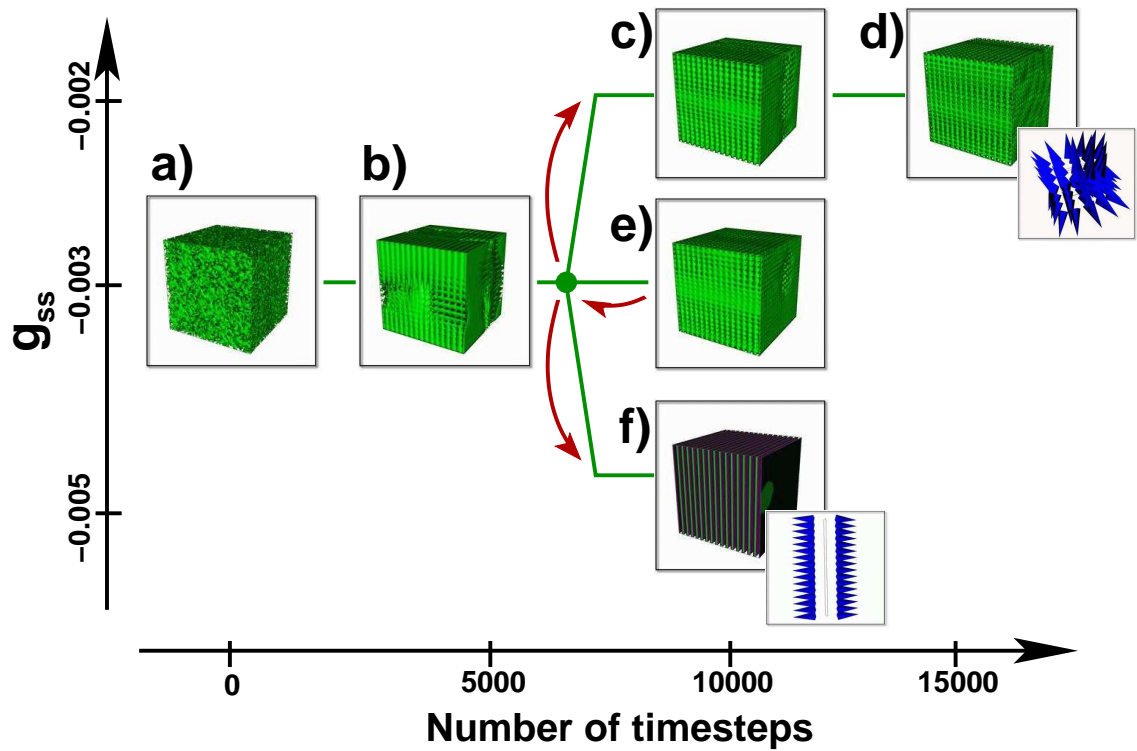}}
\caption{A steered parameter search is performed for a 64$^3$ mixture of water
and surfactant (see text for simulation details). The data visualised in the
insets (a-f) is the volume rendered surfactant density, i.e. areas with
densities higher than the average are coloured in green.
We start at timestep zero with a random fluid
mixture and $g_{ss}$ set to -0.003 (a). After 10000 timesteps we find a stable
phase, i.e. a large fraction of the surfactant molecules forms spherical
micelles (e). We `rewind' the simulation to timestep 5000 (b) and change
$g_{ss}$ to -0.002. The system's state at timestep 10000 is only slightly
different from (e), but still evolving in time. At timestep 15000 the micelles
are much clearer than at timestep 10000, showing that a larger fraction of the
surfactant molecules is involved in the micelle formation (d).
The inset in (d) shows the orientations of surfactant particles around an area of
low surfactant concentration: all of them point to regions of higher
water concentration. We now rewind to timestep 5000 again and change $g_{ss}$
to -0.005. At timestep 10000 the system has formed a lamellar structure (f),
where the surfactant molecules are aligned in parallel bilayers
between areas of high water density (see inset of (f)). 
}
\label{steersearch}
\end{figure}

Figure \ref{steersearch} depicts a steered parameter search using the LB3D
code. Instead of trying to cover the full parameter space, we only perform
a limited number of steered simulations.  We start with a random
water-surfactant mixture with the surfactant-surfactant coupling constant
$g_{ss}$ set to -0.003 and the water-surfactant coupling constant $g_{bs}$
set to -0.006 (figure \ref{steersearch}a). The lattice size is 64$^3$ and
initial maximum occupation numbers in this case are 0.7 for surfactant and
0.4 for water.  The insets of figure \ref{steersearch} show volume
rendered surfactant densities, where densities higher than the average are
coloured in green. We monitor the state of the system while the
simulation evolves and find a stable micellar phase after about 10000
timesteps (figure \ref{steersearch}e), i.e. a large fraction of surfactant
molecules forms spherical micelles.

Since we are not able to detect any drastic changes for later timesteps, we
`rewind' the simulation back to timestep 5000 and change $g_{ss}$ to -0.002.
In this way we lower the interaction between surfactant molecules. Rewinding
to an earlier simulation step is necessary because due to the low fluid
velocities it takes a very long time in a system that is close to equilibrium
for parameter changes to take effect. Steering the values of the coupling
constants more drastically is not a solution to this problem because the
simulation might become unstable and produce unphysical results or even crash.

Five thousand timesteps after `rewinding', the system arrives in a state that
is only slightly different from the $g_{ss}$=-0.003 case at timestep 10000. In
order to investigate whether any more changes occur, we let the simulation
evolve for a further 5000 timesteps and discover more well defined structures
than before. A closeup of the dataset (see inset of figure \ref{steersearch}d)
allows us to investigate the self-assembled spherical micelles in our system in
more detail: the surfactant molecules visualised by the blue cones are pointing
away from regions of low surfactant density, that is away from regions of high
water density. The more well defined structure than in figure \ref{steersearch}c is due to a larger fraction of surfactant molecules being involved in forming micelles.

In order to investigate areas of the parameter space where the absolute
value of $g_{ss}$ is higher, we `rewind' the simulation again and change
$g_{ss}$ to -0.005. At timestep 10000, the system now finds itself in a
drastically different state from the previous cases: water and
surfactant form lamellae, the surfactant molecules align in parallel
bilayers between areas filled with water. The colour purple is used
here to visualise the interface between water and surfactant and the inset
of figure \ref{steersearch}f depicts how the surfactant molecules are
aligned in this case.
 
Of course, a single steered simulation cannot by itself replace a full
task farm parameter search, but a small number of steered simulations
can provide a coarse grained overview of the available parameter space. By
steering into areas of interest, one is able to dramatically reduce the
resources required.  Most of the analysis takes place during the
simulation time itself and therefore required additional effort for
offline analysis is reduced. Moreover, CPU time is further reduced because
not every simulation has to start from timestep zero again. In the example
given here, three conventional simulations of 15000 timesteps each sum up
to 45000 simulation steps in total. In contrast, the steered simulations
use only about 25000 timesteps because we do not have to rewind to
timestep zero and can stop as soon as we cannot detect any further
changes. Since the scientist interacts with the simulations, data dumping
rates can be adapted during the run, thus reducing the requirements for
disk storage even further.

\section{Computational steering}
As alluded to in previous  sections, the problems associated with high
performance   computational  science  in   general  and   large  scale
simulations in particular are not confined to merely finding resources
with larger numbers of processors or memory.  Simulations that require
greater    computational   resources    also    require   increasingly
sophisticated and complex tools for the analysis and management of the
output of  the simulations.  In  Sections \ref{Sec:Practicalities} and
\ref{Sec:Implementation}, we  highlighted the limitations  of a simple
simulate-then-analyse  approach and  indicated how  more sophisticated
approaches help alleviate some of the problems.  We then sketched some
specific  advantages of  controlling the  evolution of  a computation,
based upon the realtime analysis  and visualisation of the status of a
simulation,  as   opposed  to  the  {\it  post   facto}  analysis  and
visualisation of the output  of a computation.  The functionality that
we  refer to  as  {\it  computational steering}  enables  the user  to
influence the  otherwise sequential simulation and  analysis phases by
merging and interspacing them.

What  additional  requirements does  computational  steering place  on
computer systems?  In order to computationally steer a simulation, one
needs an  interface to communicate  with the simulation, which  may be
running on a remote machine.  In addition to allowing parameters to be
monitored and  changed, this interface needs to  offer the possibility
of  visualising complex data  sets, for  instance 3D  isosurfacing and
volume rendering.  To enable  intuitive interaction with a simulation,
it is  essential that visualisation can be  perfomed sufficiently fast
compared to changes taking  place in the simulation.  Visualisation of
large  and  complex data  sets  typically  requires high-end  graphics
hardware,  which, like  high-end  computing resources,  is not  always
available locally.   Therefore, visualisation  should be treated  as a
distributed resource as  well, the need for which  stems not only from
computational  steering  but  also   from  the  requirements  of  high
performance  visualisation.   The requirement  to  use  more than  one
distributed resource simultaneously in  turn raises more subtle issues
associated   with  the   requirements   of  sophisticated   scheduling
algorithms  and techniques.   Typical supercomputer  centres currently
make no provision for coallocation of resources, for example a compute
node and  a visualisation node.  It  would be desirable to  be able to
request resources for a  computational steering session in advance and
be assured of a certain  quality of service during a session.  Equally
important is the requirement to be able to reserve substantial compute
resources with small turn around time.

What  advantages does computational  steering provide  the application
scientist in return? We have  described a few specific examples of how
computational steering can  increase a scientist's productivity.  This
increase in  productivity is due to  an increase in  the throughput of
hardware resources but equally  important is the enhanced productivity
due  to   a  more  effective  computational   science  workflow  bench
(simulation-analysis  loop) as  a  consequence of  being  able to  use
computational steering.  There have been attempts to use computational
steering  as a novel  approach to  studying outstanding  and important
problems  in biomolecular  systems ~\cite{SMD1}.   At the  very least,
computational  steering  complements existing  techniques~\cite{SMD2}.
However, computational steering should not  simply be thought of as an
effective tool in  the production and analysis phase  of a simulation.
It can  provide the application scientist (often  also the application
developer) with greater flexibility  in the development, debugging and
validation  phases   of  an  application~\cite{vetter95,  Liere:1997a,
Wijk:1997},  where  it complements  rather  than  replaces other  well
established  tools.  Computational  steering can  also be  extended to
collaborative  environments where  several geographically  distributed
scientists  can simultanously  interact with  one or  more simulations
from   separate  locations   ~\cite{gridbook2,   gridbook6,  brodlie1,
brodlie2, Leeds:Iris}.

This sets  the stage for a  few remarks on what  kind of computational
science  applications   are  suitable   candidates  for  the   use  of
computational  steering.   If an  application  requires  barely a  few
seconds  of   computing  time  to  simulate  a   physical  process  or
effectively  finish  the  simulation  (say  a fixed  small  number  of
iterations  in a minimization  routine), then  the advantages  of user
intervention  while the  simulation is  in progress  are limited.  Any
overhead associated  with interrupting such  an inexpensive simulation
will not be worth the gain  bought by interactivity.  At the other end
of the  spectrum, simulations that `take forever'  for any discernible
changes  to manifest as a consequence of  user interaction are also not
good candidates as the  advantage from such interactivity is typically
limited.   A case  in point  are {\it{ab  initio}}  quantum mechanical
molecular dynamics simulations, where even when only a small number of
atoms are of  interest, each step of the  calculation may take several
hours  on  a multiprocessor  machine~\cite{Harting:2000,Mishima:1998}.
Any  changes initiated dynamically  by the  user for  such simulations
would take many hours to  become manifest, clearly limiting their use.
Thus it appears  to be the case that simulations with  a run time from
several minutes to several hours (irrespective of the resources used),
are ideally suited for  interactive aspects of computational steering.
It is  important to distinguish between  the role of  steering in long
running simulations as opposed to simulations with long response times
to a pertubation. We have discussed  the limited role of steering in the
latter, but  in the former case,  steering can be  useful for checking
the  progress  of a  long  running  simulation  by connecting  to  the
simulation, getting a sample and visualising it and then disconnecting
after  checking all  is correct,  thus enabling  the scientist  to use
computational  steering as  `simulation  monitor' and  as a  safeguard
against possible wastage of computational resources.

A few cautionary remarks are in place. Many physical systems have long
equilibration time scales and suffer from finite size effects. In such
cases,  changing  the  parameters  and taking  the  state  immediately
following  the  change  to  be   the  putative  true  state  might  be
misleading.  Moreover, many  physical systems exhibit hysteresis, that
is their properties at a  given point in parameter space are dependent
upon their history.  Before  computationally steering a simulation, it
is imperative to  determine if it displays hysteresis  and, if so, how
the use of steering may influence the analysis. Computational steering
of diverse applications may involve different challenges, but in all
cases an antidote to possible  problems will be careful and consistent
study rather than a refusal to adopt new analysis techniques.

We end  this section  by discussing why  there has  been comparatively
limited  acceptance or  use  of computational  steering in  scientific
applications until  now.  If there is  a lesson to be  learnt from the
evolution of computational science, it is that the complexity of doing
something new and exciting has  to be well hidden from the application
scientist,  i.e.    it  is  essential   to  minimize  the   amount  of
implementation,   learning,  disruption  and   changes  to   the  user
interface,  until  the  advantages   of  the  new  features  are  well
established   and  very   clearly   seen  to   offset   the  cost   of
implementation. Ideally one  would like, if possible, to  just slip in
the functionality where the user  never knows or notices, but this may
not  be  achievable in  a  computational  steering  context, when  the
application   scientist   often   has   to  actually   execute   the
functionality.   Thus maximal  effort will  have to  be  invested into
reducing   the   `barrier   of   entry'.   Until   now,   implementing
computational steering  has required  a high degree  of customization,
but most scientists  typically are not in a position  to invest in the
time-consuming  task  of   developing  the  necessary  tools;  indeed,
cooperation   with  specialists   in  visualisation   and  interfacing
techniques has hitherto been vital.  This problem is not helped by the
fact  that the  requirements  of  a scientist  might  change during  a
project   because   new   results   from   simulations   lead   future
investigations  in different directions  than initially  planned. This
could result in the steering tools  having to be adapted, which can be
very costly~\cite{Liere:1997a,Wijk:1997}.

The question that logically follows is what can be done to address the
relatively low  acceptance of computational  steering in computational
science.  We  believe that  most scientists are  not cognisant  of the
advantages  computational steering  offers and  thus unaware  how they
might   benefit   from   steerable  applications.   Therefore,   their
simulations  are done in  the `traditional'  way, invoking  long batch
jobs and subsequently lengthy offline data analysis.  In this paper we
have outlined  the advantages that computational  steering has brought
to  our  LB studies.   Part  of  the purpose  of  this  article is  to
encourage computational scientists to think about the enhanced ability
and benefits  that computational steering capabilities  would bring to
their scientific productivity, along the  lines of our LB studies, but
specific  to their  own  applications.   It is  obvious  that not  all
computational  science problems are  amenable to,  or for  that matter
require, computational steering.  However, we believe that documenting
the   advantages  of  computational   steering  in   widely  differing
applications and areas will  help bring greater acceptance of steering
as a valid paradigm for  computational science research.  We also wish
to emphasize that, as shown  by the RealityGrid steering framework, by
using   the  correct  abstractions   and  good   software  engineering
practices, implementing the required  changes is much less effort than
might  otherwise be  expected.  Indeed  many  generic tools  and
libraries useful  in program steering  and data visualisation  are now
readily available ~\cite{vetter95,Beazley:1996,Prins:1999,gridlab}.

\section{Steering on Computational Grids: Current Status and Future Outlook}
\label{Sec:Outlook}

Significant  effort  is being  invested  worldwide  in Grid  computing
~\cite{ggf}.   A basic  premise of  grid computing  is to  provide the
infrastructure  required to  facilitate the  collaborative  sharing of
resources.   The grid  aims to  present  the elements  required for  a
computational task  (e.g.  calculation engine,  filters, visualisation
capability) as  components which can be  effectively and transparently
coupled through the grid framework.  In this scenario, any application
or simulation code can be  viewed simply as a data producing/consuming
object on the  grid and computational steering is a  way of allowing a
user to interact with such objects. As discussed in previous sections,
a scientist using steering has heterogeneous and dynamic
computational resource requirements, making  the stated ability of the
grid to  collectively and transparently marshall diverse resources  complementary to
the  primary  motivation of  computational  steering.     Thus   a grid
infrastructure  that  permits the  coordination  of heterogeneous  and
distributed computing resources provides a natural environment as well
as  a  testbed for  demonstrating the  effectiveness  of  steering  in
computational science.

In our  description of the  RealityGrid steering framework  in section
\ref{Sec:Implementation},   we  did   not   mention  the   use  of   a
computational  grid   or  dependence  on   any  underlying  middleware
requirement.  This  is because  the RealityGrid steering  framework is
capable of being used on stand  alone workstations as well as the most
ambitious computational  grids available  in the world  today. Equally
important is  the fact that  our steering framework is  not critically
dependent on any one particular  middleware although it does adhere to
the best  practises and the  open standards currently  being discussed
within the  Global Grid Forum  (GGF)~\cite{ggf}.  In the  remainder of
this section  we will  describe how the  RealityGrid project  uses the
grid to implement computational steering.

In the same  way that a high-fidelity simulation  of a physical system
often  requires   a  supercomputer,   so  the  visualisation   of  the
(potentially)  large data  sets  that these  simulations produce  also
requires specialist  hardware that  few scientists have  direct access
to.   Consequently,  the visualisation  component  of the  RealityGrid
steering may well be on a  machine other than the one the scientist is
sitting in front of. This is consistent with our earlier proposal that
visualisation  be  treated like  a  distributed  resource.  This  then
requires that  the output (images)  of the visualisation  component be
returned to  the user's workstation  quickly enough to allow  for full
interactivity (e.g.   to rotate, zoom, etc.).  We currently  use SGI's
OpenGL VizServer  software to perform  this task; it takes  the images
directly  from the  rendering hardware  on the  visualisation machine,
compresses them,  transports them to the  user's machine, decompresses
and  displays  them ~\cite{VizServ}.   This  allows  the scientist  to
interact with a remote visualisation, even over network links with
relatively low bandwidth.

The outline  traced above  has been the  basis for  several successful
computational  steering demonstrations that  we have  performed within
the  past year. In  our first  demonstration at  the UK  e-Science All
Hands  Conference  in  Sheffield  in  September 2002,  we  used  this
technique to interact with a visualisation produced on an SGI Onyx300 in
Manchester from  a laptop in Sheffield with  the computation performed
in London.   For this  demonstration and the  subsequent two,  we used
Unicore~\cite{unicore} as the underlying middleware to manage the file
transfer  aspects of  the demonstration.   At Supercomputing  2002, we
used a  trans-Atlantic link  to interact with  a visualisation  on the
Onyx300 in Manchester  from a laptop in Baltimore,  USA, the computation
being performed on the SGI  Origin 3800 in Manchester.  In February of
2003 at the SGI VizSummit, we  used a laptop in Paris to interact with
simulations  on  128 processors  of  the  Origin  3800 in  Manchester,
visualising  and  steering being  performed  locally  using the  Onyx300
facilities provided on  the demonstration floor.  Thus we  have in the
process performed computational  steering using three different albeit
transient   grid  scenarios:   within  the   UK,   trans-Atlantic  and
UK-continental Europe.

The grids used in these demonstrations were assembled especially for
each  event. However, the  UK e-Science  community has  constructed an
ambitious  Level 2  Grid ~\cite{l2g}  that  aims to  provide the  user
community with a  persistent grid.
We have  already deployed  a preliminary RealityGrid  LB3D application
involving computational steering, using Globus (as opposed to Unicore,
confirming the flexibility of our  steering framework) on this Level 2
Grid,  thus being amongst the  first groups  in the
world  to use  a persistent  grid for  routine science  requiring high
performance computing and computational steering.

At  the  time  of  writing,  the  RealityGrid  steering library  supports  both
file-based and streaming (based on \textit{globus\_io} from the Globus
project~\cite{globus})  data  transfer  between  the  application  and
visualisation  components. Communications between the application and the
client is currently implemented by exchanging XML documents
through a shared file system (XML is a widely accepted language specifying the syntax to mark-up data in computer documents). 

\begin{figure}
\centerline{\includegraphics[width=9.5cm,clip,angle=0.0]{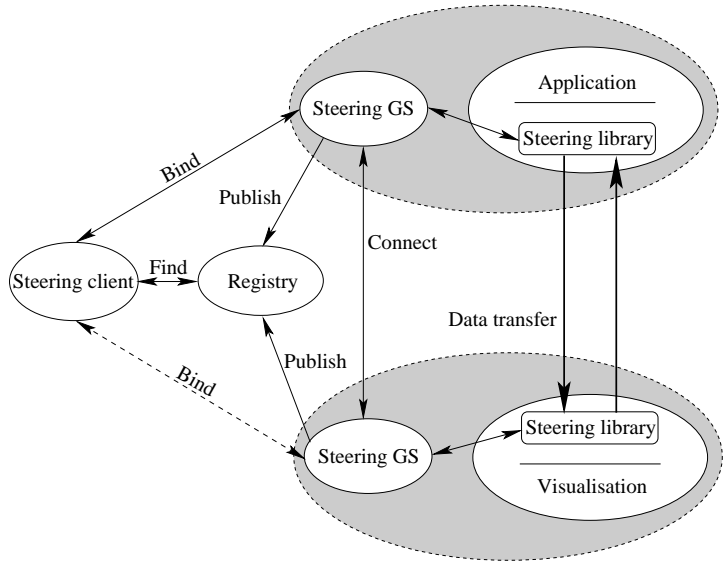}}
\caption{Architecture  for RealityGrid steering  within the  Open Grid Services
Infrastructure (OGSI). The application and client communicate by exchanging
messages through intermediate grid services. The grid service (GS) provides
the public interface  through which  clients can steer the applications.}
\label{fig:ogsa_arch}
\end{figure}

We are in the process of implementing a more flexible architecture,
based on the Open  Grid  Services Infrastructure~\cite{OGSI}.
As shown in Figure~\ref{fig:ogsa_arch}, communications between the
application and client are routed through an intermediate steering
grid service (SGS). The SGS provides the public interface through which
clients can steer the application. In our architecture, the visualisation
and application components appear on equal footing, and a visualisation can
possess its own SGS. Each SGS publishes information about
itself in a registry, which is used by clients to discover and
bind to running applications, and can also be instrumental in
bootstrapping the communications between the application and
visualisation components. We note that the approach of exposing
steering controls as grid services in a standard way could bring
profound benefits in the form of interoperability between different
implementations of computational steering.

\label{Sec:Imp:CurrentTech} 

\section{Conclusions}
\label{Sec:Conclusion}
This  paper  has  described  the  work we  have  done  to  incorporate
computational steering  in mesoscale lattice  Boltzmann simulations of
binary and ternary immiscible  and amphiphillic fluids.  The scale and
efficiency of these  studies is set to increase  dramatically with the
advent of computational grids  which are now becoming widely available
within the UK, Europe, the USA and the Pacific Rim.

\section*{Acknowledgements}
We are grateful to ESPRC for funding much of this research through RealityGrid
grant GR/R67699 and for providing access  to SGI Origin 3800, Origin 2000, and
CRAY  T3E supercomputers at Computer Services for Academic Research (CSAR),
Manchester, UK.  We  would also  like to thank the University of Manchester for
access to their SGI Onyx300 and HEFCE for funding the 16 processor dual pipe
SGI Onyx2 at University College London. Jonathan Chin acknowledges Huntsman   and  Queen Mary, University of London,
for   funding his Ph.D.  studentship. Jens Harting  wishes to   thank  the  European
Commission   Access  to  Research   Infrastructures action   of  the `Improving
Human Potential  Programme' for  supporting his stay at Italy's national
supercomputer centre in Bologna (CINECA) and the use of their local IBM SP4,
SGI Origin 3800 and SGI Onyx2 facilities.

\bibliographystyle{abbrv-unsrt} 
\bibliography{main}
 
\newpage
Jonathan Chin  is a postgraduate student  in Peter Coveney's  group at the
University  College  London.  In   1999,  while  an  undergraduate  at  the
University of Oxford,  he took a summer placement  in the group, where
he wrote  the LB3D parallel lattice  Boltzmann code and  enjoyed it so
much he  joined the group in 2000  to do a PhD.  His interests include
complex fluids, visualization, and premature micro-optimization.\\

Jens Harting studied  physics at the Carl von  Ossietzky University in
Oldenburg,   Germany.  After   a  Diploma   thesis   on  Bose-Einstein
Condensation he worked on path integral Monte Carlo simulations of few
electron systems such as semiconductor  quantum dots and received his PhD
in  December   2001.  Being  interested  in   simulations  of  fluids,
computational steering and high performance computing, joined  the
RealityGrid  project  as a  Research Fellow  in Peter Coveney's group in
March 2002.\\

Shantenu Jha is  a Research Fellow with RealityGrid  at the Centre for
Compuatatioal Science,  UCL, London. His  graduate work is  in Physics and
Computer Science from  Syracuse University, New York, USA.  His interests
are in computational  physics, high performance  and distributed computing
and politics. \\

Peter Coveney holds a Chair in Physical Chemistry in the Department of
Chemistry  and is  Director of  the Centre  for  Computational Science
(CCS) at University College London. His group performs internationally
leading research in the area  of atomistic and mesoscale modelling and
simulation,   including  molecular   dynamics,   dissipative  particle
dynamics,  lattice-gas and  lattice-Boltzmann techniques  and exploits
state of the art high performance computing and visualisation methods.
He has published  numerous theoretical and modelling/simulation papers
on lattice-gas and lattice-Boltzmann  automata, dissipative particle  dynamics and molecular
dynamics {\em inter alia}. Professor Coveney is currently leading the large RealityGrid
research  programme,  funded  by  the UK's  Engineering  and  Physical
Science Research Council, aimed  at grid enablement of supercomputing,
visualisation           and           computational           steering
(http://www.realitygrid.org). He previously held the Chair in Physical
Chemistry in the Department of  Chemistry at Queen Mary, University of
London, before which he  was with the Schlumberger Cambridge Research,
where he held a number of scientific and management positions.\\

Andrew  Porter   is  a   software  engineer  in   the  Supercomputing,
Visualization  and  e-Science group  in  Manchester  Computing at  the
University  of Manchester. He  graduated with  a PhD  in computational
condensed-matter physics from the University of Cambridge in 2000 and,
after a spell as an  IT consultant in industry, joined the RealityGrid
project in  March 2002.  His interests include  computational steering
and scientific visualisation.\\

Stephen Pickles is Software Infrastructure Manager for the RealityGrid
project, and co-leader of Manchester Computing's e-Science team at the
University of Manchester. He has  been engaged in grid computing since
1999.  After nearly  a decade as  programmer and systems  analyst with
ICL (Australia),  Stephen graduated from Macquarie  University in 1994
with BSc (Hons  I) in Physics, gained his PhD in  lattice quantum
chromodynamics from the
University  of Edinburgh  in 1998 and  then  joined the  CSAR service  at
Manchester Computing as Senior Applications Consultant.

\end{document}